\documentclass[prb]{revtex4}
\usepackage{times}
\usepackage{graphicx}
\usepackage{psfrag} 
\usepackage{fancyhdr}
\begin{document}
\date{\today}
\title{To grow or to shrink: A tale of two 
rubber balloons}
\author{\bf Yan Levin and Fernando L. da Silveira} 
\affiliation{\it Instituto de F\'{\i}sica, Universidade Federal
do Rio Grande do Sul\\ Caixa Postal 15051, CEP 91501-970, 
Porto Alegre, RS, Brazil\\ 
{\small levin@if.ufrgs.br}}

\begin{abstract}

Two identical rubber balloons are partially inflated with air (to different
extent) and connected by a hose with a valve.  
It is found that depending on balloon volumes, 
when the valve is opened the air
will flow  either from the larger (fuller) balloon to the
smaller (emptier) balloon, or from the smaller balloon 
to the larger one.  The phenomenon is explained in terms 
of the non-ideal rubber elasticity of balloons. The full phase 
diagram for the air flow dynamics is constructed.
\end{abstract}
\pacs{64.10.+h, 05.70.-a}
\maketitle
\bigskip
%\narrowtext
%\pagestyle{fancy}
%\lhead{\thepage}
%\rhead{To grow or to shrink?}
%\cfoot{}

\section{Introduction}

Consider two identical rubber balloons, the same one as
used at children's parties, filled with air and 
connected by a hose with a valve. For simplicity,
we shall approximate the balloons by spheres.
Suppose that the larger balloon, the one that
has been inflated more,
has  radius  $R^b$ and contains $n_b$ moles of air, while 
the smaller one has radius $R^s$ and contains $n_s$ moles of 
air. When the valve is
opened which balloon is going to grow and which
will shrink? It seems almost obvious that the 
larger balloon should shrink, while the smaller one 
should grow.  This, however, is not
what is often found. For some balloon sizes the air
flows from the larger balloon to the smaller one, but for other
sizes the direction of the air flow is reversed.   
What can explain this seemingly counterintuitive
behavior? 
 
\section{Thermodynamics}

When the valve of the hose connecting the two balloons
is opened, the direction of air flow is determined by
the two laws of  thermodynamics~\cite{LaLi80}.  
Suppose that $dn=-dn_b=dn_s$ moles of air  are transfered from the 
bigger balloon to the
smaller one. The work needed to accomplish this is determined by  
the first law of thermodynamics, 
%-----------------------------
\begin{equation}
\label{1} 
dW=T dS_0+ \sum_{i=s,b} (dE_i+dE_i^r+P_0 dV_i) \;,
\end{equation}
%-----------------------------
where $S_0$, $P_0$ and $T$ are the  environment entropy, pressure,
and temperature; $E_i$, $P_i$, and $V_i$ are the 
internal energy, pressure and  volume
of the gas inside the two balloons  $i=b,s$; and   $dE_i^r$
is the internal energy of the balloon rubber membranes.
The second law of thermodynamics requires that
%-----------------------------
\begin{equation}
\label{1aa} 
dS_0+dS_b+dS_s+dS_b^r+dS_s^r \ge 0 \;,
\end{equation}
%-----------------------------
where $S_i$ and $S_i^r$ are the entropy of gas 
and of the rubber membrane  of balloon $i$.

Since the transfer of $dn$ moles of gas between the two
balloons with finite  pressure difference is an irreversible process,
the total entropy of the universe will increase.  It follows that the
work necessary for the transfer is bounded from below by
%-----------------------------
\begin{equation}
\label{1ab} 
dW > \sum_{i=s,b} (dE_i+dE_i^r-T dS_i-T dS_i^r+P_0 dV_i) \;.
\end{equation}
%-----------------------------
From thermodynamics
%-----------------------------
\begin{equation}
\label{1a} 
dE_i=T dS_i-P_i dV_i + \mu_i dn_i \;,
\end{equation}
%----------------------------- 
and
%-----------------------------
\begin{equation}
\label{1ac} 
dE_i^r=T dS_i^r-P_i dV_i^r +\sigma_i dA_i \;,
\end{equation}
%----------------------------- 
where $V_i^r$ is the volume of rubber,  $\sigma_i$ is the surface
tension, $A_i$ is the surface area, and $\mu_i$ is the
chemical potential of gas inside the balloon $i$.
 
As a leading order approximation we can consider rubber to be 
incompressible, so that $dV_i^r=0$. 
Substituting Eqs. \ref{1a} and \ref{1ac} into Eq.~(\ref{1ab}), we obtain
%-----------------------------
\begin{equation}
\label{1b} 
dW>\sum_{i=s,b} [(P_0-P_i) dV_i+\sigma_i dA_i+\mu_i dn_i]\;.
\end{equation}
%-----------------------------
For a spherical balloon,  the variation in the  volume and
in the surface area are related by
%-----------------------------
\begin{equation}
\label{2} 
\frac{dV_i}{dA_i}=\frac{R_i}{2} \;,
\end{equation}
%----------------------------- 
while the difference between the internal
and the  external pressures is governed by the law of Laplace~\cite{LaLi80},
%-----------------------------
\begin{equation}
\label{3} 
P_i-P_0=\frac{2 \sigma}{R_i} \;.
\end{equation}
%-----------------------------
Substituting  Eqs.~(\ref{2}) and (\ref{3}) into Eq. (\ref{1b}),
the amount of work necessary to accomplish the transfer of air is
%-----------------------------
\begin{equation}
\label{4} 
dW>(\mu_s-\mu_b) dn \;.
\end{equation}
%-----------------------------
%We could have arrived at Eq.~(\ref{4}) quicker by noting that
%work that must be done on a system in contact with a thermal reservoir and
%kept at constant external pressure $P_0$ is ,
%-----------------------------
%\begin{equation}
%\label{5} 
%dW \ge dG\;,
%\end{equation}
%----------------------------- 
%where the Gibbs free energy of the system is
%-----------------------------
%\begin{equation}
%\label{6} 
%G=\mu_b n_b + \mu_s n_s\;.
%\end{equation}
%-----------------------------
%The equality holds only for reversible processes.
%Substituting Eq.~(\ref{6}) into Eq.~(\ref{5}) we arrive directly at
%Eq.(\ref{4}).

For gas at fixed temperature, the Gibbs-Duhem equation~\cite{Ch87} is
%-----------------------------
\begin{equation}
\label{7} 
-V_i dP_i+n_i d\mu_i=0 \;,
\end{equation}
%-----------------------------
so that 
%-----------------------------
\begin{equation}
\label{8} 
\left. \frac{\partial \mu_i}{\partial P_i}\right|_T=\frac{V_i}{n_i}  \;.
\end{equation}
%-----------------------------
Using the ideal gas equation of state and integrating, the 
chemical potential
inside each balloon is found to be, 
%-----------------------------
\begin{equation}
\label{9} 
\mu_i=\mu_0+R T \ln\left(\frac{P_i}{P_0}\right) \;,
\end{equation}
%-----------------------------
where $\mu_0$ is the reference chemical potential at atmospheric
pressure.  Inserting Eq.~(\ref{9})
into Eq.~(\ref{4}),  the amount of work 
necessary to transfer $dn$ moles of gas from the larger balloon
to the smaller one is,
%-----------------------------
\begin{equation}
\label{10} 
dW>R T \ln\left(\frac{P_s}{P_b}\right) dn \;.
\end{equation}
%-----------------------------

Transfer of gas will occur spontaneously if it does not
require any external work, $dW=0$. 
For the specific case of two interconnected balloons 
this  will be the case
if the flow of air is from  balloon with high internal 
pressure to the one with low internal pressure. Furthermore, 
since the thermodynamics
requires that 
$\partial \mu/\partial P>0$ see Eq.~(\ref{8}), this conclusion
holds for non-ideal gases as well.

Although the calculation was performed 
for spherical balloons, it is evident that the 
conclusion concerning the direction of air flow,  Eqs.~(\ref{4})
and (\ref{10}),
does not depend on the specific balloon shape. 
%While the final result --- air flows from 
%the balloon with high internal pressure to the one with low 
%internal pressure ---
%seems self-evident, proving this from first principles
%is not trivial.  Furthermore, this
%conclusion is not correct in general.   
On the other hand, if  the same balloons 
are filled with a heavy gas and are placed at
different heights in the gravitational field, the flow
of air will no longer be governed simply by the difference in internal
pressure between the two balloons.  Nevertheless, while Eq.~(\ref{10})
will loose its validity,  Eq.~(\ref{4}) will remain correct, however
the chemical potential of gas  will have to include a gravitational
contribution as well.

\section{Rubber elasticity}

As air enters into  
balloon, its rubber  membrane stretches and
becomes tense. The Helmholtz free energy of a rubber
sheet is
%-----------------------------
\begin{equation}
\label{11} 
F=E-T S\;.
\end{equation}
%-----------------------------
If the sheet is subjected to  stress $\sigma$, 
its internal energy $E$ will 
vary in accordance with Eq.~(\ref{1ac}),
and its Helmholtz free energy as,
%-----------------------------
\begin{equation}
\label{11b} 
dF=dE-T dS-S dT=-P dV-S dT+\sigma dA\;.
\end{equation}
%-----------------------------
The surface tension of the sheet is then,
%-----------------------------
\begin{equation}
\label{11c} 
\sigma=\left .\frac{\partial F}{\partial A} \right |_{V,T}=
\left .\frac{\partial E}{\partial A} \right |_{V,T}-
\left . T \frac{\partial S}{\partial A} \right |_{V,T} \;.
\end{equation}
%-----------------------------
It is an experimental fact~\cite{Flo53} that up to fairly 
large extensions the internal energy of a rubber sheet
is independent of its area,
%-----------------------------
\begin{equation}
\label{11d} 
\left .\frac{\partial E}{\partial A} \right |_{V,T}=0 \;.
\end{equation}
%-----------------------------
This equation should be compared to a similar equation for the ideal
gas ---  the internal energy of an ideal gas is independent of the
volume that it occupies. Clearly, this is not correct for real gases, but 
serves as a very good approximation for a gas at low density.  Similarly,
Eq.~\ref{11d} holds only for ``ideal'' rubber.

The Flory theory~\cite{Flo53} 
allows us to calculate the change in entropy of an
elastic object during a deformation.  Suppose that the 
rubber sheet has dimensions, $L_x$, $L_y$, and $L_z$ and that after the
deformation the new dimensions are $\lambda_x L_x$, $\lambda_y L_y$, and 
$\lambda_z L_z$, then
%-----------------------------
\begin{equation}
\label{11e} 
\Delta S= -k [\lambda_x^2+\lambda_y^2 +\lambda_z^2-3-
\ln(\lambda_x\lambda_y\lambda_z)] \;,
\end{equation}
%-----------------------------
where $k$ is a constant related to the number of chains and
the topological structure
of the polymer network.  If  rubber is subjected to a not
very high stress it is reasonable  to assume that it is incompressible,
so that $\lambda_x \lambda_y \lambda_z=1$. This means that for
a uniform isotropic stress in the $x - y$ plane, 
$\lambda_x =\lambda_y \equiv \lambda$ and $\lambda_z=1/\lambda^2$.
Eq.~(\ref{11e}) then  simplifies to
%-----------------------------
\begin{equation}
\label{11f} 
\Delta S= -k \left [ \frac{2 A}{A_0} +\frac{A_0^2}{A^2}-3\right] \;,
\end{equation}
%-----------------------------
where $A_0$ is the surface area prior to the deformation and $A$ is the
final surface area. 
It is important to note that a 
deformation of an elastic body results in a decrease of its
entropy.  Microscopically, this is a consequence of the 
reduction of the conformational volume accessible to a stretched polymer.
Substituting Eq.~(\ref{11f}) into Eq.~(\ref{11c}) 
the surface tension of a rubber sheet is
%-----------------------------
\begin{equation}
\label{11g} 
\sigma = \kappa \left [ 1-\frac{ A_0^3}{A^3}\right] \;,
\end{equation}
%-----------------------------
where $\kappa=2 k T/A_0$. 

We expect Eq.~(\ref{11g}) to work reasonably well up to extensions
on the order of $100 \%$.  
If the balloon is inflated beyond this ``ideal''
limit,  deviations are to be expected. 
%%%%%%%%%%%%%%%% figure %%%%%%%%%%%%%%%%%%%%%
%\begin{figure}[h]
%\begin{center}
%\psfrag{R_s}{ $R^*$}
%\includegraphics[width=6cm]{fig1.eps}
%\end{center}
%\vspace{1cm}
%\caption{Stylized curve of the surface tension vs. surface area 
%of an inflating balloon. }
%\label{fig1}
%\end{figure}
%%%%%%%%%%%%% end of figure %%%%%%%%%%%%%%%%% 
In general then,
the stress-strain relation for a rubber balloon of radius $R$ 
can be written as
%-----------------------------
\begin{equation}
\label{12} 
\sigma(R,R_0) \simeq \kappa \left(1-\frac{R_0^6}{R^6}\right) 
f\left(\frac{R}{R^*}\right)\;,
\end{equation}
%-----------------------------
where $R_0$ is the balloon radius prior to inflation, and
the scaling function $f(x)$ governs the crossover from the
ideal rubber regime to the non-ideal one.  The scale
$R^*$ specifies the  balloon size at which its rubber starts
to behave non-ideally.  Here we 
use the simplest form of the crossover function,
%-----------------------------
\begin{equation}
\label{13} 
 f(x) = \left\{
\begin{array}{ll} 
1 & {\rm for}\;\;  0 \le x \le 1, \\
A (x-1)^\alpha +1&  {\rm for}\;\; x>1
\end{array}  
\right.
\end{equation}
%-----------------------------
where $A>0$ and  $\alpha>1$. Note that the 
function $f(x)$ is continuous and
differentiable everywhere.
  
Transfer of $dn$ moles of air from the  bigger balloon
to the smaller one will occur spontaneously if
the pressure inside the bigger  balloon is larger than the
pressure inside the smaller balloon. To simplify the notation
lets measure all the lengths in units of $R_0$ and define
$R^s=u R^b$ with $\{1/R^b \le u \le 1\}$. 
Using the law of Laplace
%-----------------------------
\begin{equation}
\label{14} 
g(u; R^b ,R^*, A) \equiv P_b-P_s=\frac{2 \sigma(R^b)}{ R^b}-\frac{2 \sigma(u R^b)}{u R^b} \;.
\end{equation}
%-----------------------------
If $g > 0$, once the valve is open 
there will be a spontaneous transfer of gas from the larger balloon
to the smaller.  On the other hand, if $g < 0$ the transfer
will proceed in the opposite direction. From Eqs.~(\ref{12})
and (\ref{13}) we see that $g(1;R^b, R^*, A)=0$ while
%-----------------------------
\begin{equation}
\label{15} 
\lim_{u \rightarrow 1/R^b } g(u; R^b, R^*, A) = \frac{2 \sigma(R^b)}{ R^b}>0\;.
\end{equation}
%-----------------------------
Therefore, unless the function $g$  has a zero
on the interval $\{1/R^b \le u \le 1\}$ the larger balloon will
always shrink, passing its air to the smaller balloon. 
Indeed for a given $f(x)$, we find that if  $R^*<R^*_m$ 
then  $g > 0$
for all  $\{1/R^b \le u \le 1\}$, and the larger balloon  always
inflates the smaller one, see Fig. 1.  The maximum 
value of the crossover scale $R^*_m$ and the balloon size $R^*_m$
for which this behavior occurs is determined by the set of equations, 
%-----------------------------
\begin{equation}
\label{16} 
g'(1; R^b_m, R^*_m, A) = 0 \;,
\end{equation}
%-----------------------------
and 
%-----------------------------
\begin{equation}
\label{17} 
g''(1; R^b_m, R^*_m, A) = 0 \;,
\end{equation}
%-----------------------------
where the prime denotes differentiation with respect to $u$.

We now distinguish two cases: $1<\alpha<2$ and  $\alpha \ge 2$. 
For $\alpha \ge 2$  and  $R^*>R^*_m$ the equation 
%-----------------------------
\begin{equation}
\label{18} 
g'(1; R^b, R^*, A) = 0 \;,
\end{equation}
%-----------------------------
has two roots $R^b_1$ and $R^b_2$ and a new phase in which
small balloons inflate the larger ones appears, see Fig. \ref{fig3}.
This phase terminates at a critical point, beyond which  a 
large balloon with size $R^b>R^b_c$ will  
inflate any smaller balloon. 
For a given crossover function and crossover scale $R^*$, 
the location of the critical point is determined by the set of
equations,
%-----------------------------
\begin{equation}
\label{18a} 
g(u_c; R^b_c, R^*, A) = 0 \;,
\end{equation}
%-----------------------------
and
%-----------------------------
\begin{equation}
\label{18b} 
g'(u_c; R^b_c, R^*, A) = 0 \;.
\end{equation}
%-----------------------------
%%%%%%%%%%%%%%%% figure %%%%%%%%%%%%%%%%%%%%%
\begin{figure}[h]
\begin{center}
\psfrag{a1}{ $\alpha> 1$ and $R^*<R^*_m$}
\psfrag{R_b}{ $R^b$}
\psfrag{R_s}{ $R^s$}
\psfrag{R_1}{ $R^b_1$}
\psfrag{R_2}{ $R^b_2$}
\psfrag{R_3}{ $R^b_3$}
\psfrag{R_c}{ $R^b_c$}
\includegraphics[width=6cm]{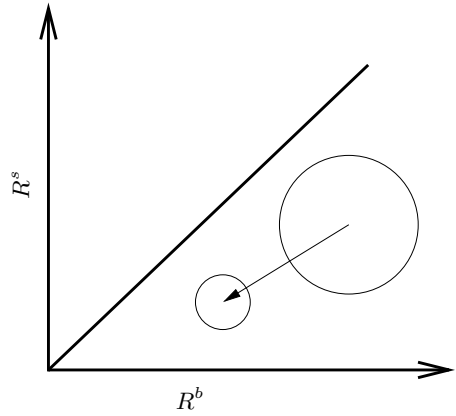}
\end{center}
%\vspace{1cm}
\caption{Phase diagram for  $R^*<R^*_m$. 
The $45^o$ line is $R^b=R^s$. 
Arrow shows the direction of
the air flux. The big balloon always inflates
the smaller one.  }
\label{fig2}
\end{figure}
%%%%%%%%%%%%% end of figure %%%%%%%%%%%%%%%%% 

%%%%%%%%%%%%%%%% figure %%%%%%%%%%%%%%%%%%%%%
\begin{figure}[h]
\begin{center}
\psfrag{a2}{ $\alpha > 1$ and $R^*>R^*_m$}
\psfrag{R_b}{ $R^b$}
\psfrag{R_s}{ $R^s$}
\psfrag{R_1}{ $R^b_1$}
\psfrag{R_2}{ $R^b_2$}
\psfrag{R_3}{ $R^b_3$}
\psfrag{R_c}{ $R^b_c$}
\includegraphics[width=6cm]{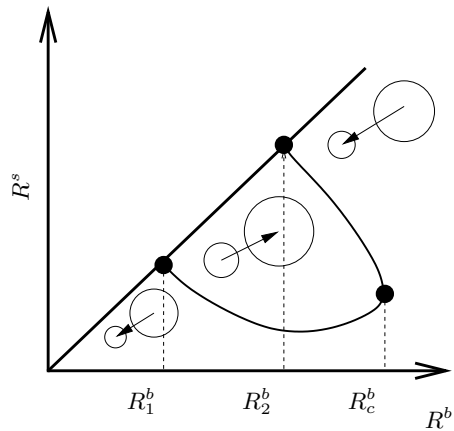}
\end{center}
%\vspace{1cm}
\caption{Phase diagrams for  $R^*>R^*_m$. 
There are two distinct regimes, separated by a 
phase boundary  terminating at a critical point, beyond which
a large balloon will inflate any  smaller balloon.}
\label{fig3}
\end{figure}
\begin{figure}[h]
\begin{center}
\psfrag{R_b}{\Large $R^b$}
\psfrag{R_s}{\Large $R^s$}
\psfrag{R_1}{\large $R^b_1$}
\psfrag{R_2}{\large $R^b_2$}
\psfrag{R_3}{\large $R^b_3$}
\psfrag{R_c}{\large $R^b_c$}
\includegraphics[width=8cm]{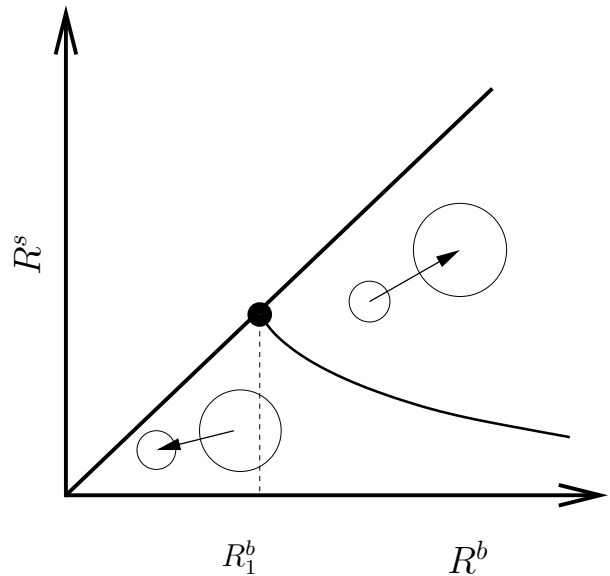}
\end{center}
%\vspace{1cm}
\caption{Phase diagram for the ideal 
rubber balloons, $R^* \rightarrow \infty$. }
\label{fig6}
\end{figure}

%%%%%%%%%%%%% end of figure %%%%%%%%%%%%%%%%% 
As the crossover scale grows, so do the values of
$R^b_2$ and $R^b_c$.  For $R^*>1$
%-----------------------------
\begin{equation}
\label{19} 
R^b_2 \simeq a R^*  \;,
\end{equation}
%-----------------------------
where $a$ is the root of equation
%-----------------------------
\begin{equation}
\label{20} 
a = \frac{ A (a-1)^\alpha+1}{A \alpha (a-1)^{\alpha-1}}  \;.
\end{equation}
%-----------------------------
If the  crossover scale goes to infinity, $R^* \rightarrow \infty$, 
so that balloon rubber always behaves ideally, then
$R^b_2 \rightarrow \infty$ and 
$R^b_c \rightarrow \infty$.  In this limit the phase diagram 
assumes the topology presented in Fig. \ref{fig6}.
%%%%%%%%%%%%%%%% figure %%%%%%%%%%%%%%%%%%%%%
%%%%%%%%%%%%% end of figure %%%%%%%%%%%%%%%%%

\begin{figure}[h]
\begin{center}
\psfrag{R^*}{\Large $R^*$}
\psfrag{A}{\Large $A$}
\includegraphics[width=8cm]{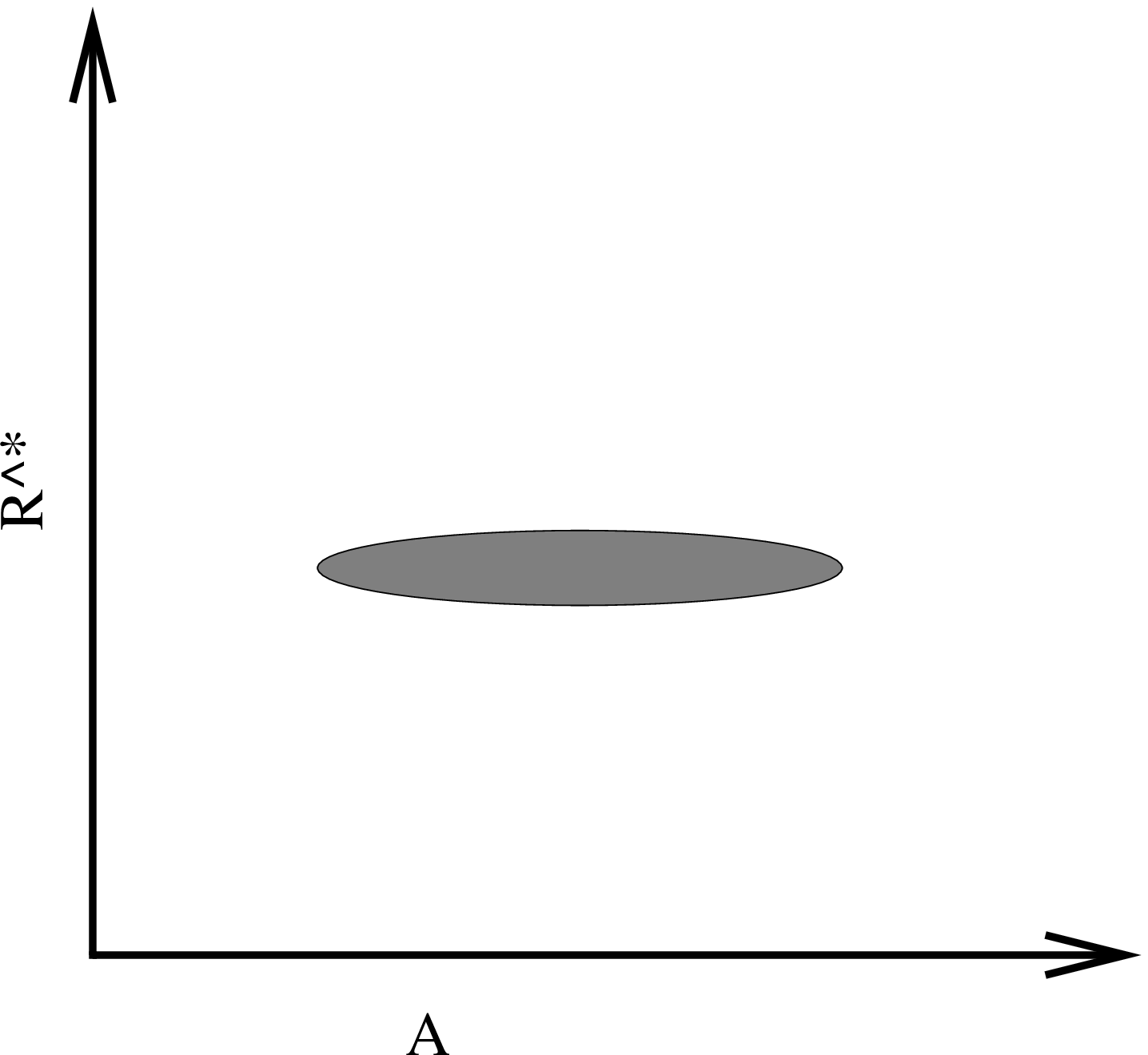}
\end{center}
%\vspace{1cm}
\caption{The shaded region indicates the parameters for which
the topology of the phase diagram is the one shown in Figs. \ref{fig4} and
\ref{fig5}.  Outside the shaded region the topology is the one
presented in Figs. \ref{fig2}, \ref{fig3}, and \ref{fig6}.  
As $\alpha \rightarrow 2$ the shaded area shrinks to zero.}
\label{fig7}
\end{figure}

\begin{figure}[h]
\begin{center}
\psfrag{R_b}{\Large $R^b$}
\psfrag{R_s}{\Large $R^s$}
\psfrag{R_1}{\large $R^b_1$}
\psfrag{R_2}{\large $R^b_2$}
\psfrag{R_3}{\large $R^b_3$}
\psfrag{R_4}{\large $R^b_4$}
\psfrag{R_c1}{\large $R^b_{c1}$}
\psfrag{R_c2}{\large $R^b_{c2}$}
\includegraphics[width=8cm]{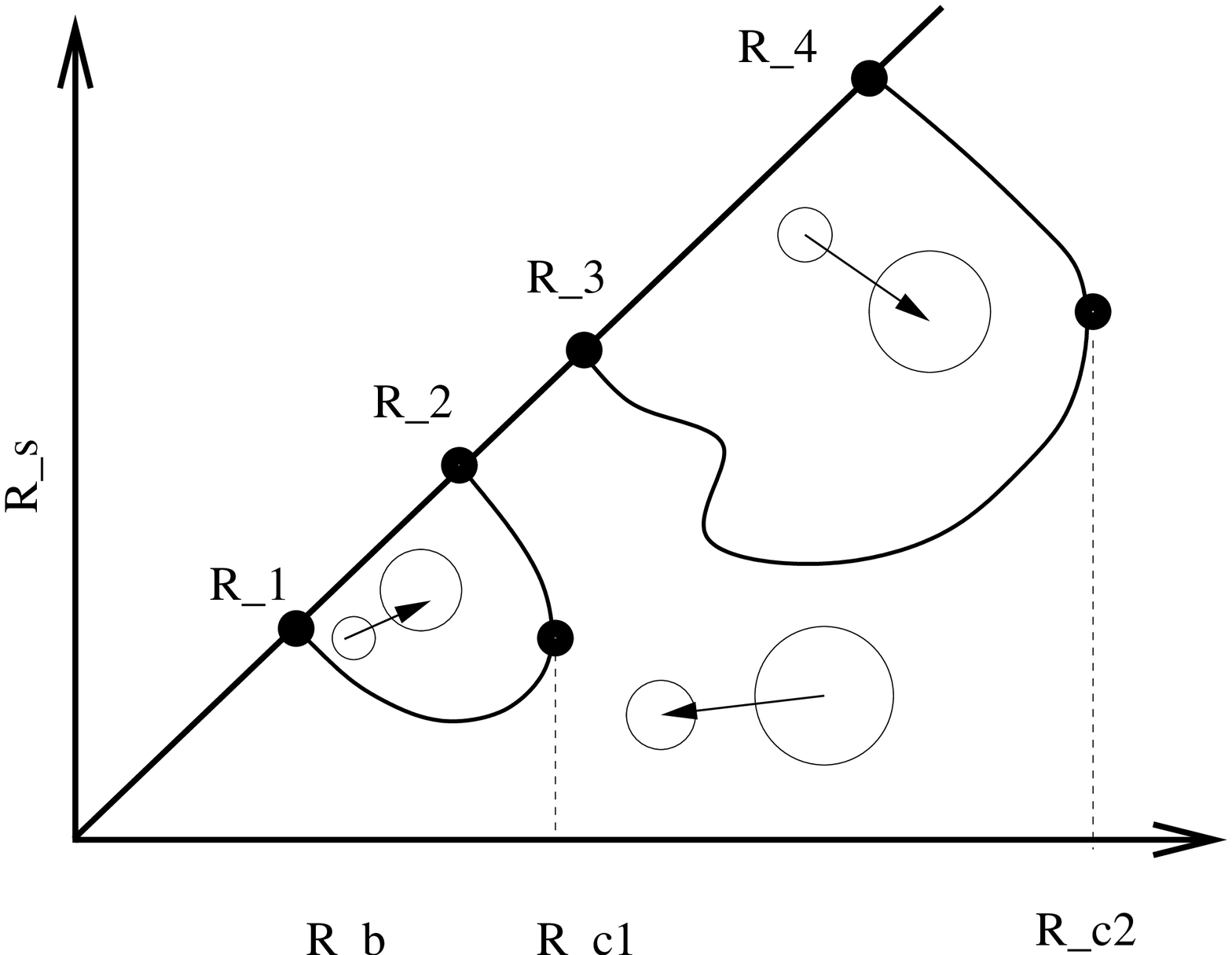}
\end{center}
%\vspace{1cm}
\caption{Possible phase diagram for $1<\alpha<2$. As 
$R^b_3 \rightarrow R^b_{c1}$ there is a topological change with
the new phase diagram shown in Fig. \ref{fig5}.}
\label{fig4}
\end{figure}

\begin{figure}[h]
\begin{center}
\psfrag{R_b}{\Large $R^b$}
\psfrag{R_s}{\Large $R^s$}
\psfrag{R_1}{\large $R^b_1$}
\psfrag{R_2}{\large $R^b_2$}
\psfrag{R_3}{\large $R^b_3$}
\psfrag{R_4}{\large $R^b_4$}
\psfrag{R_c1}{\large $R^b_{c1}$}
\psfrag{R_c2}{\large $R^b_{c2}$}
\includegraphics[width=8cm]{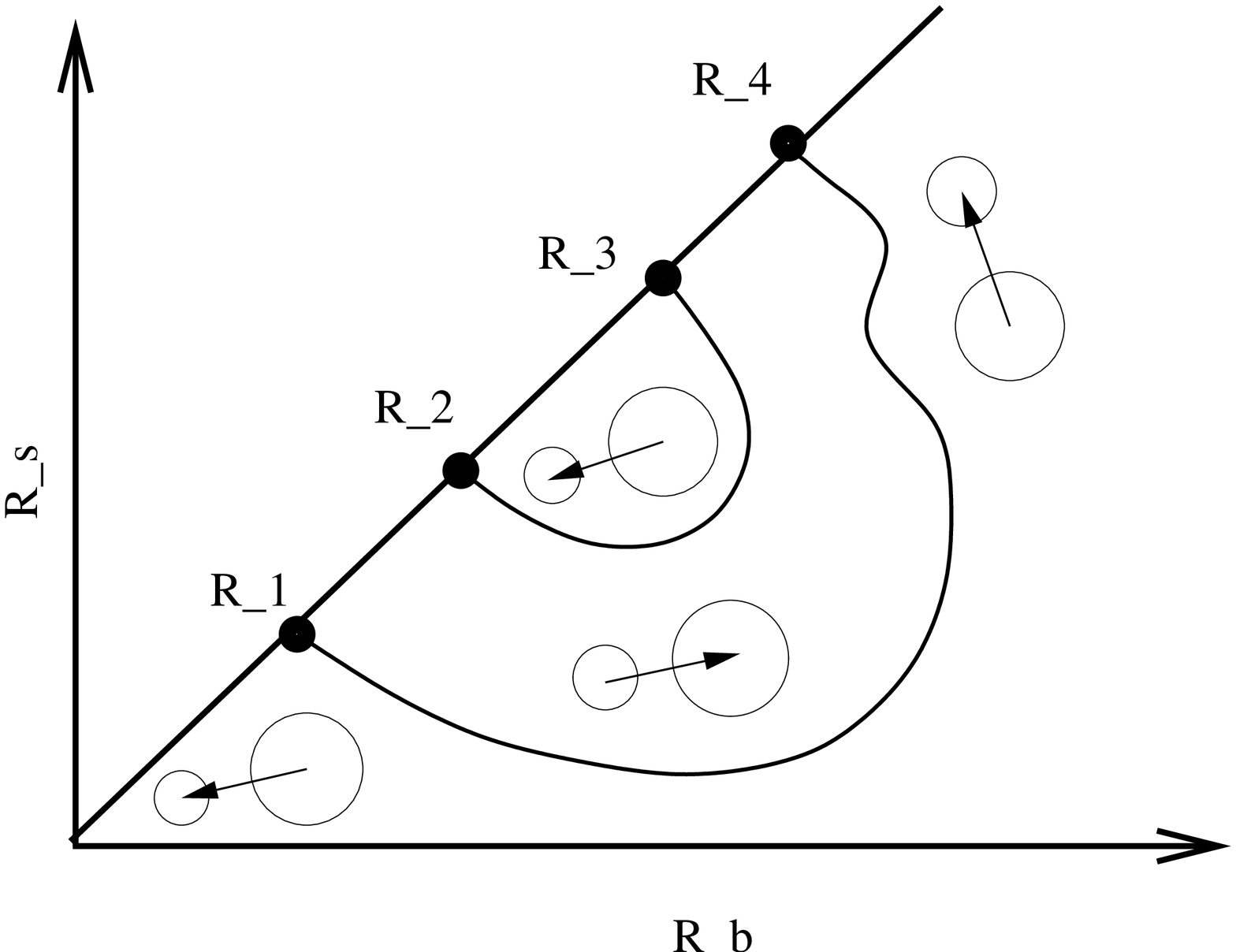}
\end{center}
%\vspace{1cm}
\caption{Possible phase diagram for $1<\alpha<2$. }
\label{fig5}
\end{figure}

For $1<\alpha<2$ the phase diagram becomes even more interesting. For
$A$ and $R^*$ inside the shaded area of Fig. \ref{fig7}, 
Eq.(\ref{18}) has
four solutions $R^b_1$, $R^b_2$, $R^b_3$, and  $R^b_4$, and the
possible topologies of the phase diagram are shown in 
Figs. \ref{fig4} and \ref{fig5}. As the boundary of 
the shaded region is approached from inside, 
one of the phases shrinks and disappears
and the phase diagram assumes the topology of Fig. \ref{fig3}. 
Outside the shaded region,
the only possible topologies are the ones presented in 
Figs. \ref{fig2}, \ref{fig3}, and \ref{fig6}.
%%%%%%%%%%%%%%%% figure %%%%%%%%%%%%%%%%%%%%%
%%%%%%%%%%%%% end of figure %%%%%%%%%%%%%%%%%
The topological change from Fig. \ref{fig4}
to Fig. \ref{fig5} occurs as $R^b_3 \rightarrow R^b_{c1}$. 
As $\alpha \rightarrow 2$, the shaded area in Fig. \ref{fig7} 
shrinks to zero and
the topology of the phase diagram reduces to the one presented in
Fig. \ref{fig3}.
%%%%%%%%%%%%%%%% figure %%%%%%%%%%%%%%%%%%%%%
%%%%%%%%%%%%% end of figure %%%%%%%%%%%%%%%%%

%%%%%%%%%%%%%%%% figure %%%%%%%%%%%%%%%%%%%%%
%%%%%%%%%%%%% end of figure %%%%%%%%%%%%%%%%%
\section{Conclusion}

We have explored the thermodynamics of air transfer 
between two partially inflated rubber balloons.
Surprisingly, for such an apparently simple system
a  very rich phase diagram governing the air transfer
between the two balloons is obtained. We find that depending
on the elasticity of balloon rubber and the 
balloon sizes, the air can flow either from the larger balloon
to the smaller one, or {\it vice versa}.  The topology of the
phase diagram is controlled by the crossover function 
which characterizes the deviation of balloon rubber from the
ideal Flory behavior.

%\section{Acknowledgements}
This work was supported in part by the Brazilian agencies
%Conselho Nacional de Desenvolvimento Cient{\'\i}fico e Tecnol{\'o}gico 
CNPq and FAPERGS.

\bigskip

%\newpage

%\bibliographystyle{plain}
%\bibliographystyle{prsty}
%\bibliography{references}

\end{document}